\documentstyle[preprint,seceq,wrapft,epsf]{ptptex}

\newcommand{\beq}{\begin{equation}}
\newcommand{\eeq}{\end{equation}}
\newcommand{\bea}{\begin{eqnarray}}
\newcommand{\eea}{\end{eqnarray}}

\newcommand{\sla}[1]{{/ \!\!\! #1}}
\newcommand{\Sla}[1]{{/ \!\!\!\! #1}}
\newcommand{\wt}[1]{{\widetilde #1}}

\newcommand{\br}[1]{{\bar #1}}
\newcommand{\cl}[1]{{\cal #1}}

\newcommand{\del}{\partial}
\newcommand{\la}{\langle}
\newcommand{\ra}{\rangle}
\newcommand{\dgr}{\dagger}

\newcommand{\tr}{{\rm tr}}

\newcommand{\Ncr}{N^\infty_{\rm cr}}
\newcommand{\VEVofMass}{\la \br{\psi} \psi \ra_\Lambda}
\newcommand{\TwoDFour}{2 \! < \! D \! < \! 4}
\newcommand{\Dim}{$D(\TwoDFour)$ }
\newcommand{\AD}{\gamma_{\br{\psi}\psi}}

\notypesetlogo  
\preprintnumber{
DPNU-96-59}
\title{
Fermion Mass Generation in the $D$-dimensional Thirring Model
 as a Gauge Theory
}

\author{
Masaki {\sc Sugiura}
}

\inst{
Department of Physics, Nagoya University, Nagoya 464-01%
}

\recdate{
\today
}

\abst{
Based on the Schwinger-Dyson (SD) equation,
 the fermion mass generation
 is further studied in
 the $D$($\TwoDFour$)-dimensional Thirring model
 as a gauge theory previously proposed.
By using a certain approximation to the kernel,
 we analytically obtained
 explicit form of the dynamical mass of fermion
 and the critical line in $(N,1/g)$ space,
 where $N$ is the number of fermions
 and $g$ is the dimensionless
 vector-type four-fermion coupling constant.
This analytical result is confirmed
 by the numerical solution
 for the SD equation with exact form of the kernel
 in (2+1) dimensions.

}

\begin{document}
\maketitle
\section{Introduction}

Through the studies
 of the dynamical origin of the quarks/leptons mass,
 it was discovered{}\cite{Hol,MY,MTY} \ 
 that the dynamics inducing the {\it large anomalous dimension}
 of the operator $\br{\psi}\psi$ is strongly required.\cite{Hol} \ 
In particular,
 the scalar/pseudoscalar-type
 four-fermion interactions with the gauge interaction
 (gauged Nambu-Jona-Lasinio model{}\cite{BLL,KMY,MY})
 have played a very important role
 as a renormalizable model in (3+1) dimensions ($D=4$).\cite{KTY} \ 
It has also been shown{}\cite{KTY} \ that
 the phase structure of such a gauged NJL model
 in (3+1) dimensions is quite similar
 to that of the $D$$(\TwoDFour)$-dimensional
 four-fermion theory of scalar/pseudoscalar-type
 (without gauge interactions),\cite{KY} \ 
 often called Gross-Neveu model,\cite{GN} \ 
 which is renormalizable in $1/N$ expansion.\cite{RWP}

Recently a model with only
 the vector-type four-fermion interaction
 has been studied as another candidate for the scenario of
 the fermion dynamical mass generation,%
\cite{HP,IKSY,Kondo,Hands,KmKm} \  namely,
 the $D$($\TwoDFour$)-dimensional Thirring model{}\cite{Thi} \ 
 with $N$ fermions.
If we naively treated this model with the usual gap equation
 made by $1/N$-leading diagram,
 we would conclude no fermion mass generation.
Nevertheless,
 we can obtain the dynamical mass of fermion
 by solving the SD equation with the rainbow diagram
 of the composite vector boson
 which is induced in $1/N$-leading order.\cite{HP} \ 
Then the dynamical mass has a nontrivial
 dependence on $1/N$ as a consequence of the SD equation
 similar to the case of QED in (2+1) dimensions.\cite{ANW} \ 

In our previous paper{}\cite{IKSY} \ 
 we stressed that
 the composite vector boson
 is actually a dynamical gauge boson
 corresponding to the hidden local symmetry (HLS){}\cite{BKY} \ 
 being broken spontaneously,
 which means that the Lagrangian can be rewritten
 into the manifest gauge symmetric form.
Combined with the nonlocal gauge fixing,\cite{GSC} \ 
 this manifest gauge degree of freedom
 plays an important role
 in construction of the SD equation
 to keep the consistency
 between the ladder approximation
 and the Ward-Takahashi identity.
Based on this SD equation,
 we performed the proof that
 the fermion dynamical mass
 is actually generated at finite value of $N$
 ($N$: number of four-component fermions)
 in \Dim dimensions,%
	\footnote{%
		The original proof in Ref.~\citen{IKSY}
		 includes the case of $D=2$,
		 although the exact solution{}\cite{Thi} \ 
		 seems to tell us no mass generation.
		In Appendix \ref{App:Bifurcation}
		 we will show
		 why this difference occurs.
	} \ %
 through the context of Ref.~\citen{MN}.
Moreover we obtained explicit form
 of the mass function and a critical value of $N$
 in the limit of $g \rightarrow \infty$
 in (2+1) dimensions,
 where $g$ is the dimensionless four-fermion coupling constant
 of the Thirring model.

Further studies were done
 by several authors.\cite{Kondo,Hands,KmKm} \ 
The author of Ref.~\citen{Kondo} derived
 the explicit form of the critical value of $N$ for finite $g$
 (denoted as $N_{\rm cr}(g)$)
 in the $D$-dimensional Thirring model
 by using the inversion method instead of the SD equation.
The explicit form of $N_{\rm cr}(g)$
 is somewhat different from the one given in Ref.~\citen{Kondo}
 and this paper (See \S\ref{sec:Anal_Sol}
	\footnote{%
		Although we give only the critical value of $g$,
		 as $g_{\rm cr}(N)$,
		 we can read $N_{\rm cr}(g)$
		 by inverting $g = g_{\rm cr}(N_{\rm cr}(g))$
		 with respect to $N_{\rm cr}(g)$.
	}%
 ),
 although they agree with each other qualitatively
 as to the existence of the critical $N$.
In Refs.~\citen{Hands,KmKm}
 the lattice simulation were performed
 in (2+1) dimensions,
 and the results seem to support
 those of analytical studies.\cite{IKSY,Kondo}

In this paper
 we first recapitulate the formalism of Ref.~\citen{IKSY}
 for the Thirring model as a gauge theory
 and the SD equation in the nonlocal gauge.
We then give a revised proof of
 the existence of a nontrivial solution,
 since the previous proof{}\cite{IKSY} \ 
 was based on an implicit assumption and thus was not complete.

Next we show explicit form of the analytical solution
 in the limit of $g \rightarrow \infty$
 in $D$($\TwoDFour$) dimensions
 not restricted to (2+1) dimensions.
By introducing a certain ansatz for the kernel
 of the SD equation,
 we further obtain explicit form of the analytical solution
 for finite value of $g$
 not restricted to $g \rightarrow \infty$.
The validity of the ansatz will be checked
 by the numerical solution
 in the case of (2+1) dimensions.
The analytical solutions for the finite coupling constant
 will provide us with the information
 for investigating the phase structure, i.e.,
 the critical line in $(N,1/g)$ plane,
 the beta function, the anomalous dimension, etc..

This paper is organized as follows.
In \S{}2
 we briefly summarize the HLS
 and nonlocal gauge fixing procedure.
In the first subsection of the \S{}3
 the review will be continued to introduce
 how to construct the SD equation,
 and in the last subsection
 we will improve the existence proof
 of a nontrivial solution of the SD equation
 as a supplement to the previous one.\cite{IKSY} \ 
In \S{}4
 the analytical calculations for the SD equation
 in \Dim dimensions
 will be done to obtain explicit form of the solutions
 used to investigate the properties of the model.
\S{}5 will be devoted to the comparison
 between the analytical and the numerical solutions
 in (2+1) dimensions.
The conclusion and discussion will be given in \S{}6.

\section{Hidden Local Symmetry}
\label{sec:HLS}
\subsection{Hidden Local Symmetry in the Thirring Model}
In this and the next sections
 we would like to review briefly the formulation
 used in Ref.~\citen{IKSY}
 to investigate the dynamical mass generation of fermion
 in the $D$-dimensional Thirring model.

Let us start with the following Lagrangian (the Thirring Model):
\beq
	\cl{L}_{\rm Thi} = 
		\sum_{a=1}^N \br{\Psi}_a i \sla{\del} \Psi_a
			-
		{G \over 2N} (\sum_{a=1}^N \br{\Psi}_a \gamma^\mu \Psi_a)^2,
	\label{eqn:Org_Lagrangian}
\eeq
 where it is supposed that
 each of Dirac gamma matrices is represented formally
 as $4 \times 4$ matrix even in \Dim dimensions,
 and the subscript $a$ runs over the {\it color} of fermions.
(Hereafter we will suppress summation symbol, $\sum_{a=1}^N$.)
It is well known that this Lagrangian can be rewritten
 with the vector auxiliary field, $\wt{A}_\mu$,
 such as
\beq
	\cl{L}_{\rm Thi}
		=
		\br{\Psi}_a i \gamma^\mu
		\left(
			\del_\mu - \textstyle{{i \over \sqrt{N}}} \wt{A}_\mu
		\right)
		\Psi_a
			+
		{1 \over 2G} \wt{A}_\mu^2,
	\label{eqn:Auxiliary} \\
\eeq
 where it is easy to see that
 $\wt{A}_\mu(x)$ is related to fermions,
\beq
	\wt{A}_\mu = 
		- {G \over \sqrt{N}} \br{\Psi}_a \gamma_\mu \Psi_a,
\eeq
 from the equation of motion.
Eq.(\ref{eqn:Auxiliary}) has no local symmetry
 associated with $\wt{A}_\mu$ due to the ``mass term''
 appearing in the last of Eq.(\ref{eqn:Auxiliary}).
Nevertheless, we can rewrite Eq.(\ref{eqn:Org_Lagrangian})
 or equivalently Eq.(\ref{eqn:Auxiliary})
 so that it actually has
 {\it the hidden local symmetry} (HLS)\cite{BKUYY,GCS,KK,BKY,IKSY} \ 
 associated with the gauge field $A_\mu$
 which is introduced through the following field redefinition:
\newcommand{\amdp}{A_\mu - \sqrt{N}\del_\mu \phi}
\begin{subeqnarray}
	\label{eqn:UnitaryGauge}
	\Psi_a(x)		&=&	u^\dgr(x)\psi_a(x) = e^{-i\phi(x)}\psi_a(x),
	\\
	\wt{A}_\mu(x)	&=&	i\sqrt{N}u^\dgr \cdot D_\mu u = \amdp,
\end{subeqnarray}
 where the field $u(x) \equiv e^{i\phi(x)}$
 can be recognized as the nonlinearly-realized basis
 of the NG boson field $\phi(x)$
 accompanying the spontaneous symmetry breaking
 of the hidden local symmetry, $U(1)_{\rm hidden}$.
Thus, using the fields defined above,
 the Lagrangian (\ref{eqn:Org_Lagrangian}) is rewritten
 into the following form,
\bea
	\cl{L}_{\rm Thi}
		&=&
		\br{\psi}_a i \Sla{D} \psi_a
			-
		{N \over 2G} (u^\dgr \cdot D_\mu u)^2,
	\nonumber \\
		&=&
		\br{\psi}_a i \gamma^\mu
		\left(
			\del_\mu - \textstyle{{i \over \sqrt{N}}} A_\mu
		\right)
		\psi_a
			+
		{1 \over 2G} (\amdp)^2,
	\label{eqn:HLS}
\eea
 in which the HLS is realized manifestly at the Lagrangian level:
\beq
	\psi_a \mapsto \psi_a^\prime = e^{i\alpha} \psi_a\;,
	\;\;
	A_\mu \mapsto A_\mu^\prime = A_\mu + \sqrt{N}\del_\mu \alpha\;,
	\;\;
	\phi \mapsto \phi^\prime = \phi + \alpha\;.
	\label{eqn:HLTransf}
\eeq
``Gauge equivalence'' between Eq.(\ref{eqn:Auxiliary})
 and (\ref{eqn:HLS}) can be shown by taking particular choice
 of the gauge transformation in Eq.(\ref{eqn:HLTransf})
 as $\alpha(x) = -\phi(x)$, called the unitary gauge.
If we further identify $\psi^\prime_a = \Psi_a$ and
 $A^\prime_\mu = \wt{A}_\mu$ under such a gauge,
 Eq.(\ref{eqn:HLTransf}) is identical
 to Eqs.(\ref{eqn:UnitaryGauge}).

Why should the existence of the HLS be emphasized?
The reasons are the following:
First, coupled with the BRS formalism,\cite{KO} \ 
 this local symmetry might enable us
 to prove the S-matrix unitarity more straightforwardly.
Secondly, actual calculations, particularly loop calculations,
 are generally hopeless in the unitary gauge,
 while the HLS provides us to take the most appropriate gauge
 for our purpose.
For these reasons,
 the BRS gauge fixing procedure should applied
 to Eq.(\ref{eqn:HLS}) instead of the unitary gauge.
Moreover, we should introduce a derivative- (momentum-) dependent
 gauge fixing parameter $\xi(\del^2)$ ($\xi(k^2)$), i.e.,
 the nonlocal gauge fixing\cite{GSC} \ 
 since it is needed to guarantee the Ward-Takahashi identity
 in the ladder SD equation (see \S\ref{sec:SD_eqn}).
It has also been proven that
 we can construct the BRS symmetric Lagrangian
 even for the nonlocal gauge.\cite{IKSY} \ %

Here we adopt the $R_{\xi}$ gauge fixing term such as
\beq
	\cl{L}_{\rm GF}
		=
	- \frac12
		\left(
			\del_\mu A^\mu + \sqrt{N}{\xi(\del^2) \over G} \phi
		\right)
		{1 \over \xi(\del^2)}
		\left(
			\del_\nu A^\nu + \sqrt{N}{\xi(\del^2) \over G} \phi
		\right),
	\label{eqn:GFFinal}
\eeq
 where $\xi(\del^2)$ is
 a nonlocal (momentum dependent) parameter.
Therefore the total Lagrangian is given by
\bea
	&&
	\cl{L}_{\rm total}
		=
	\cl{L}_{\rm Thi} + \cl{L}_{\rm GF}
		=
	\cl{L}_{\psi,A} + \cl{L}_{\phi},
	\\
	&&
	\cl{L}_{\psi,A} = 
		\br{\psi}_a i \Sla{D} \psi_a
			+
		{1 \over 2G}(A_\mu)^2
			-
		\frac12 \del_\mu A^\mu
		\left(
			{1 \over \xi(\del^2)} \del_\nu A^\nu
		\right),
	\label{eqn:TotalLag}\\
	&&
	\cl{L}_{\phi} = 
		\frac12 \del_\mu \phi \del^\mu \phi
			-
		{1 \over 2G} (\xi(\del^2)\phi)\phi,
	\label{eqn:DecouplingLag}
\eea
 where the $\phi$ was rescaled
 as $\sqrt{N/G}\phi \mapsto \phi$.
In this gauge fixing
 the fictitious NG boson $\phi$ is completely decoupled
 independently whether $\xi$ is nonlocal or not.
Equation (\ref{eqn:TotalLag}) appears
 as if we added the ``covariant gauge fixing term''
 to Eq.(\ref{eqn:Auxiliary}).
However, without introducing the HLS,
 there is no reason why we should add such a term.
Such a confusion was actually made
 by some authors{}\cite{GMRS,HP} \ 
 who happened to arrive at the Lagrangian
 having the same form as Eq.(\ref{eqn:Auxiliary})
 in the case of constant $\xi$.
Here we stress again that
 this Lagrangian (\ref{eqn:TotalLag}),
 whether the gauge parameter is nonlocal or not,
 can only be justified through the HLS in the $R_{\xi}$ gauge.

\section{Schwinger-Dyson Equation and Dynamical Mass Generation}
\label{sec:SD_eqn}
The purposes of this section are
 to briefly review how to construct the SD equation
 of the $D$-dimensional Thirring Model{}\cite{IKSY}\ 
 and to improve the existence proof of a nontrivial solution
 of the SD equation given in Ref.~\citen{IKSY}.

\subsection{Schwinger-Dyson Equation in the nonlocal $R_\xi$ gauge}
The SD equation for the fermion full propagator
 $S(p) = i[A(-p^2)\sla{p} - B(-p^2)]^{-1}$,
 with $B(p)$ being the order parameter
 of the chiral symmetry breaking,
 is written as follows:
\bea
	&&
	(A(-p^2) - 1) \sla{p} - B(-p^2)
	\nonumber \\
	&&\mbox{\hspace{3em}}
		= - \frac1N \int {d^D q \over i(2 \pi)^D}
			\gamma_{\mu}
			{
				A(-q^2) \sla{q} + B(-q^2)
			\over
				A^2(-q^2) q^{2} - B^{2}(-q^2)
			}
			\Gamma_{\nu}(p,q)\;
			i D^{\mu\nu}(p-q),
	\label{eqn:SDeqn}
\eea
 where $\Gamma_{\nu}(p,q)$ and
 $D_{\mu\nu}(p-q)$ denote the full vertex function and
 the full propagator of the dynamical gauge boson,
 respectively.
As indicated in our previous paper,\cite{IKSY} \ 
 the R.H.S. of Eq.(\ref{eqn:SDeqn}) is $O(1/N)$,
 which implies that there is no nontrivial solution
 in the $1/N$ expansion.
Thus we should solve it and find the solution
 which depends on $1/N$ in a non-analytic way
 similar to the case of the QED$_3$.\cite{ANW}

We should apply some appropriate approximations
 to Eq.(\ref{eqn:SDeqn})
 so as to reduce it to the soluble integral equation
 for the mass function $M(-p^2) = B(-p^2)/A(-p^2)$.
Approximations adopted here are the following:
First, bare vertex (ladder) approximation is used
 to $\Gamma^\mu(p,q)$, i.e., $\Gamma^\mu(p,q) = \gamma^\mu$.
Secondly, only the one-loop diagram
 of the bare fermion propagators $i/\sla{p}$ is taken
 to evaluate the vacuum polarization tensor $\Pi_{\mu\nu}(-k^2)$.
This approximation may be consistent
 with both $A(-p^2) = 1$ in the nonlocal gauge fixing
 and the bifurcation theory used in later sections,
 which reduces the SD equation as a nonlinear integral equation
 to the linear one with respect to $B(-p^2)$.

The vacuum polarization tensor in these approximations
 is given by
\bea
	\Pi_{\mu\nu}(k)
		&\equiv&
	\left(
		g_{\mu\nu} - {k_\mu k_\nu \over k^2}
	\right)
	\Pi(-k^2),
	\label{eqn:VPT1} \\
	\Pi(-k^2)
		&=&
	-C_D^{-1} \cdot (-k^2)^{{D-2 \over 2}},\;\;
	C_D^{-1}
		\equiv
	{2 \tr I \over (4\pi)^{D/2}}
	\Gamma( 2 - \textstyle{{D \over 2}} )
	B( \textstyle{ {D \over 2} , {D \over 2} } ),
	\label{eqn:VPT2}
\eea
 where $\tr I$,
 equal to four by an assumption
 argued in \S{}\ref{sec:HLS},
 is the trace of unit matrix in spinor indices,
 and thus the gauge boson full propagator is as follows:
\bea
	iD_{\mu\nu}(k)
		&=&
	id(-k^2)
	\left[
		g_{\mu\nu} - \eta(-k^2){k_\mu k_\nu \over k^2}
	\right],
	\label{eqn:Propagator} \\
	d(-k^2)
		&=&
	{1 \over G^{-1} - \Pi(-k^2)},\;\;
	\eta(-k^2)
		=
	{
		\xi(-k^2)\Pi(-k^2) - k^2
	\over
		\xi(-k^2)G^{-1} - k^2
	}.
	\label{eqn:GaugeFixFunc}
\eea
Then the SD equation (\ref{eqn:SDeqn}) is reduced
 to the following coupled equations for $A(-p^2)$ and $B(-p^2)$:
\newcommand{\inner}{\!\cdot\!}
\begin{subeqnarray}
	\label{eqn:SeparatedSDeqn}
	A(-p^2) - 1
		\!\!&=&\!\!
	{1 \over N p^2} \int {d^D q \over i(2 \pi)^D}
	{A(-q^2) \over A^2(-q^2) q^2 - B^2(-q^2)}
	\nonumber \\
		& &
	\hspace{-4em}\times \;
	d(-k^2)
	\left[
		\{ \eta(-k^2) + 2 - D \} (p \inner q)
			- { 2 (k \inner p)(k \inner q) \over k^2 } \eta(-k^2)
	\right],
	\slabel{eqn:SeparatedSDeqn_a}
	\\
	B(-p^2)
		\!\!&=&\!\!
	- \frac1N \int {d^D q \over i(2 \pi)^D}
	{B(-q^2) \over A^2(-q^2) q^2 - B^2(-q^2)}
	\nonumber \\
		& &
	\hspace{10em} \times \;
	d(-k^2) [D-\eta(-k^2)],
	\slabel{eqn:SeparatedSDeqn_b}
\end{subeqnarray}
 where $k_\mu = p_\mu - q_\mu$.

It is generally difficult to deal with the coupled equations.
We follow the nonlocal gauge
 proposed by Georgi et al.\cite{GSC,KEIT} \ 
 which reduces the coupled SD equations
 into a single equation for $B(-p^2)$.
Due to the freedom of gauge choice of $\xi(-k^2)$,
 the R.H.S. of Eq.(\ref{eqn:SeparatedSDeqn_a}) can be
 set to zero, which implies $A(-p^2) \equiv 1$.
In such a nonlocal gauge $B(-p^2)$ itself is a mass function,
 i.e., $M(-p^2)= B(-p^2)$.

Requiring $A(-p^2) \equiv 1$ in the nonlocal gauge,
 Eqs.(\ref{eqn:SeparatedSDeqn}) are translated
 into the following forms
 (hereafter the Euclidean notation is used):
\begin{subeqnarray}
	\label{eqn:LadderSDeqn}
	0
	\hspace{-.5em}
		&=&
	\hspace{-.5em}
	\int_0^{\pi} \! d\theta \sin^D \theta
	\bigg[
        {1 \over D-1} {d \over d k^2}
		\left\{
			d(k^2) (\eta(k^2) + D - 2)
		\right\}
	\nonumber \\
		&&
	\hspace{16em}
		+
   	    {\eta(k^2) d(k^2) \over k^2}
	\bigg],
	\slabel{eqn:LadderSDeqn_a} \\
	B(p^2)
	\hspace{-.5em}
		&=&
	\hspace{-.5em}
	\frac1N \int_0^{\Lambda^{D-2}} \!\!
	d(q^{D-2}) K(p,q;G)
	{q^2 B(q^2) \over q^2+B^2(q^2)},
	\slabel{eqn:LadderSDeqn_b}
\end{subeqnarray}
 where the kernel $K(p,q;G)$ is given by
\begin{subeqnarray}
	\label{eqn:SDKernel}
	K(p,q;G)
		&=&
	{
		1
	\over
		2^{D-1} \pi^{{(D+1) \over 2}}
		(D-2) \Gamma({D-1 \over 2})
	}
	\int^\pi_0 d \theta \sin^{D-2} \theta
	\nonumber \\
		&&
	\hspace{8em}
	\times
	d(k^2) [D-\eta(k^2)].
	\slabel{eqn:Form_Kernel} \\
	\big(
		k^2 \!\!
			&=&
		\!\! p^2 + q^2 - 2pq \cos\theta.
	\big)
	\slabel{eqn:KernelArg}
\end{subeqnarray}
The kernel Eq.(\ref{eqn:Form_Kernel}) has the following properties;
 symmetry under the exchange among first two arguments
 as $K(p,q;G) = K(q,p;G)$,
 positivity and finiteness summarized as
 $0 \!<\! K(p,q;G) \!<\! \infty$ for $G \! > \! 0$.
Here the ultraviolet momentum cut-off $\Lambda$ is needed
 since the integral of the R.H.S. of Eq.(\ref{eqn:LadderSDeqn_b})
 diverges according to the power counting.
Now the role of the equations (\ref{eqn:LadderSDeqn})
 appears to be separated.
Eq.(\ref{eqn:LadderSDeqn_a}) determines the momentum dependence
 of $\eta(k^2)$ ($\xi(k^2)$) as
\beq
	\eta(k^2)
		=
	-
	{
		(D-2)
	\over
		k^{2(D-1)}d(k^2)
	}
	\int^{k^2}_0 d\zeta \zeta^{D-1} d^\prime(\zeta),
	\label{eqn:Eta_ExplicitForm}
\eeq
 which is obtained by integrating the R.H.S.
 with respect to $k^2$.
Eq.(\ref{eqn:LadderSDeqn_b}) is therefore reduced
 to the integral equation with respect to the only one function,
 $B(p^2)$.

\subsection{Existence Proof of a Nontrivial Solution %
 and the Critical Line}
\label{sec:Proof_of_NontrivialSolution}
{}From now on we reconsider the existence proof
 of a nontrivial solution
 of the SD equation (\ref{eqn:LadderSDeqn_b}),
 based on the bifurcation theory discussed in Refs.~\citen{MN,Atk}
 and briefly summarized in Appendix \ref{App:Bifurcation}.

Eq.(\ref{eqn:LadderSDeqn_b})
 always has a trivial solution $B(p^2) \equiv 0$.
We are often interested in the vicinity
 of the phase transition point
 where the nontrivial solution also starts to exist without gap.
Such a bifurcation point is identified by
 the existence of an infinitesimal solution
 $\delta B(p^2)$ around the trivial one $B(p^2) \equiv 0$.\cite{Atk} \ 
Then we obtain the linearized equation for $\delta B(p^2)$:
\beq
	\delta B(p^2)
		=
	\frac1N \int_{m^{D-2}}^{\Lambda^{D-2}}
	d(q^{D-2}) K(p,q;G) \delta B(q^2),
	\label{eqn:Bifurcation}
\eeq
 where the IR cut-off $m$ is introduced.
It is enough for us to show the existence
 of a nontrivial solution of the bifurcation equation
 (\ref{eqn:Bifurcation}).\cite{MN} \ 
Particularly, we can obtain the exact phase
 transition point where the bifurcation takes place.
Since the solution is normalized as $m = \delta B(m^2)$,
 $m$ is nothing but the dynamically generated fermion mass.

Rescaling $p = \Lambda x^{\frac{1}{D-2}}$
 and $\delta B(p^2) = \Lambda \Sigma (x)$,
 Eq.(\ref{eqn:Bifurcation}) can be rewritten as follows:
\beq
	\Sigma(x)
		=
	\frac1N \int_{\sigma_m}^{1} dy \wt{K}(x,y;g) \Sigma(y),
	\label{eqn:R_Bifurcation}
\eeq
 where $\sigma_m \equiv (m/\Lambda)^{D-2}$ ($\sigma_m < 1$),
 the dimensionless four-fermion coupling constant
 $g \equiv G\Lambda^{D-2}$ and the dimensionless kernel
 is defined by
\beq
	\wt{K}(x,y;g) \equiv
	K(x^{\frac{1}{D-2}},y^{\frac{1}{D-2}};g).
\eeq
As we mentioned in the end of the previous subsection,
 the kernel $\wt{K}(x,y;g)$ is positive, symmetric and finite:
\beq
	0 <
	\wt{K}(x,y;g)
		=
	\wt{K}(y,x;g) < \infty,\;\;\;
	\mbox{for $x$, $y$ $\in$ $[\sigma_m,1]$, $g \ge 0$}.
	\label{eqn:Property_of_K}
\eeq
These are the most important properties
 for the existence proof of a nontrivial solution.\cite{MN} \ 

Here let us consider the linear integral equation
 as an eigenvalue problem:
\beq
	\phi(x)
		=
	{1 \over \lambda}
	\int_{\sigma_m}^1 dy \wt{K}(x,y;g) \phi(y),
	\label{eqn:EigenValueProblem}
\eeq
 whose eigenvalues and eigenfunctions are denoted
 by $\lambda_n (g,\sigma_m)$
 ($| \lambda_n | \geq | \lambda_{n+1} |;\; n = 1,2,\ldots$.)
 and  $\phi_n (x)$, respectively.
The kernel $\wt{K}(x,y;g)$ is a symmetric one
 and hence satisfies the following property:
\beq
	\sum_{n=1}^{\infty} \lambda^2_n (g,\sigma_m)
		=
	\int_{\sigma_m}^1 \int_{\sigma_m}^1
	dx dy [ \wt{K}(x,y;g) ]^2 < \infty.
	\label{eqn:eigen}
\eeq
The R.H.S. of Eq.(\ref{eqn:eigen}) gives
 the upper bound for each eigenvalues $\lambda_n (g,\sigma_m)$,
 so these are finite.
Furthermore, using the positivity of the symmetric kernel
 (see Eq.(\ref{eqn:Property_of_K})),
 it is easily proven that the maximal eigenvalue
 $\lambda_1 (g,\sigma_m)$ is always positive
 and the corresponding eigenfunction $\phi_1(x)$
 has a definite sign (nodeless solution).
Moreover, we can also show that
 $\lambda_1(g,\sigma_m)$ is monotonically decreasing function
 for $\sigma_m$, i.e.,
 $\lambda_1(g,\sigma_m) < \lambda_1(g,\sigma_m^\prime)$
 for $\sigma_m > \sigma_m^\prime$,
 as proven in Appendix \ref{App:EigenValue}.
This behavior of the eigenvalue for $\sigma_m$
 was implicitly assumed in the previous proof.\cite{IKSY} \ 

In the bifurcation equation (\ref{eqn:R_Bifurcation})
 this implies the following:
 {\em If $N$ is smaller than the maximal eigenvalue
 of the kernel at $\sigma_m = 0$} : 
 {\em $N \le \lambda_1 (g,0)$},
 {\em then there exists a value of $\sigma_m$
 at which $N = \lambda_1(g,\sigma_m)$
 and the corresponding nontrivial nodeless solution
 $\Sigma(x) = \phi_1 (x)$ besides a trivial one.}
The above statement means that the parameter $\sigma_m$
 corresponds to the dynamically generated mass
 $m=\Lambda(\sigma_m)^{\frac{1}{D-2}}$.
Moreover, $N_{\rm cr}(g)$ introduced as
\beq
	N_{\rm cr}(g) = \lambda_1 (g,\sigma_m = 0)
\eeq
 determines the critical line in $(N,1/g)$ plane
 and separating it by two parts
 which are corresponding to the broken and the symmetric phase,
 since there is no non-zero solution
 for $N > N_{\rm cr}(g)$ by definition.
Existence of the critical line,
 $N=N_{\rm cr}(g)$ or $g=g_{\rm cr}(N)$,
 in the two-parameter space is somewhat analogous
 to that in the gauged NJL model.\cite{KMY}

Finally, we can point out something more about the critical line
 by using the explicit form of the kernel.
{}From Eq.(\ref{eqn:SDKernel})
 it can be seen that the kernel vanishes when $g = 0$.
This means that $\lambda_1(g=0,\sigma_m) = 0$,
 therefore the critical line starts from $(0,\infty)$.
On the other hand,
 the kernel remains finite
 when taking $g \rightarrow \infty$
 owing to the IR cut-off,
 which concludes that
 the critical line ends at $(N_{\rm cr}(\infty),0)$
 with non-zero $N_{\rm cr}(\infty)$.%
	\footnote{%
		At this stage we cannot yet conclude that
		 $N_{\rm cr}(\infty)$ is finite
		 since $\int^1_0\int^1_0dxdy\wt{K}(x,y;\infty)^2=\infty$.
		However, the explicit calculation in the next section
		 tells us that $N_{\rm cr}(\infty)$ is actually finite.
	}

\section{Analytical Solutions for $D(2 \!<\! D \!<\! 4)$ Dimensions}
\label{sec:Anal_Sol}
Without solving the SD equation (\ref{eqn:LadderSDeqn}) explicitly,
 it has been proven that
 there exists the critical line in $(N,1/g)$ plane,
 which starts from $(0,\infty)$ to $(N_{\rm cr}(\infty),0)$.
However,
 we have to find an explicit form of the solution
 to know more about the properties of the model,
 which are, for example,
 momentum dependence of the mass function,
 the beta function and the anomalous dimension
 of the operator $\br{\psi}\psi$, etc..
Therefore we attempt to solve the SD equation
 only for the particular regions of the coupling constant,
 since the integral for the kernel, Eq.(\ref{eqn:Form_Kernel}),
 cannot be performed for general value of $g$.
Furthermore we restrict ourselves
 to analyze the bifurcation equation (\ref{eqn:Bifurcation})
 instead of Eq.(\ref{eqn:LadderSDeqn_b}),
 since it is difficult to solve Eq.(\ref{eqn:LadderSDeqn_b})
 being a nonlinear integral equation.
Thus we will investigate the properties of the model
 near the critical line in $(N,1/g)$ plane,
 in which we are most interested.

\subsection{Analytical Solution {\rm :} $g\rightarrow \infty$ case}
\label{sec:Anal_Sol_1}
First we consider that $g$ is infinity,
 which is identical to the vanishing limit
 of the dynamical gauge boson mass
 proportional to $1/g$.
In this case the spontaneously broken $U(1)_{\rm hidden}$
 becomes a manifest symmetry of physical states,
 therefore we expect that the result is similar
 to that of $D$-dimensional QED,
 which we will discuss in the end of this subsection.

The functions of the propagator of $A_\mu$
 defined in Eq.(\ref{eqn:Propagator}) have now the following form.
\bea
	d(k^2)
		&=&
	{1 \over -\Pi(k^2)} = C_D k^{2-D},
	\\
	\eta(k^2)
		&=&
	{(D-2)^2 \over D},
	\label{eqn:Eta_case1}
\eea
 where $\eta(k^2)$ is determined by Eq.(\ref{eqn:Eta_ExplicitForm})
 and is independent of the momentum $k^2$,
 which means that the gauge fixing parameter $\xi(k^2)$
 is actually {\it nonlocal}
 according to Eq.(\ref{eqn:GaugeFixFunc}):
\beq
	\xi(k^2)
		=
	- \textstyle{{(D-1)(D-4) \over D}} C_D k^{4-D}.
\eeq
Furthermore,
 the integral kernel of the SD equation Eq.(\ref{eqn:SDKernel})
 can be calculated exactly as
\bea
	K(p,q;\infty)
		&=&
	K_D
	\left(
		{2 \over p+q + |p-q|}
	\right)^{D-2}
	\nonumber \\
		&=&
	K_D
	\left[
		\theta(p^{D-2}-q^{D-2}){1 \over p^{D-2}} + 
		\theta(q^{D-2}-p^{D-2}){1 \over q^{D-2}}
	\right],
	\label{eqn:Kernel_case1}\\
	K_D
		&\equiv&
	{
		(D-1) \Gamma(D)
	\over
		D (D-2) \Gamma(2-\textstyle{{D \over 2}})
		(\Gamma(\textstyle{{D \over 2}}))^3
	}.
	\label{eqn:Kernel_Coefficient}
\eea
By using the same rescaling as Eq.(\ref{eqn:R_Bifurcation}),
 such as $x = (p/\Lambda)^{D-2}$, $y = (q/\Lambda)^{D-2}$,
 $\sigma_m = (m/\Lambda)^{D-2}$ and $\Sigma(x) = B(p^2)/\Lambda$,
 we can rewrite the SD equation as
\beq
	\Sigma(x)	=
		{K_D \over N} \int^1_{\sigma_m} dy
		\left[
			\theta(x-y) {1 \over x} + \theta(y-x) {1 \over y}
		\right]
		\Sigma(y).
	\label{eqn:SDeqn_case1}
\eeq
As usually done for the SD equation of QED,
 the solution of this integral equation is obtained
 by translating Eq.(\ref{eqn:SDeqn_case1})
 into the following differential equation
 with the boundary conditions and the normalization condition:
\beq
	{d \over dx}
		\left(
			x^2 {d \Sigma(x) \over dx}
		\right)
		+ {\Ncr \over 4N} \Sigma(x)
		= 0,
	\label{eqn:Diff_Eq_case1}
\eeq
\vspace{-1em}
\begin{subeqnarray}
	\label{eqn:BoundaryCondition_case1}
	&\hbox to 8em{$\Sigma^\prime(\sigma_m) = 0,$\hfil}&
	\mbox{(IR B.C.)}
	\slabel{eqn:IRBC_case1} \\
	&\hbox to 8em{%
	$
	\left(
		x \Sigma
	\right)^\prime |_{x=1} = 0,$\hfil}&
	\mbox{(UV B.C.)}
	\slabel{eqn:UVBC_case1} \\
	&\hbox to 8em{%
	$\Sigma(\sigma_m) = \sigma_m^{1/(D-2)},$\hfil}&
	\mbox{(Normalization Condition)}
	\slabel{eqn:Normalization_case1}
\end{subeqnarray}
 where,
\beq
	\Ncr \equiv 4 K_D = 
		{
			4 (D-1)\Gamma(D)
		\over
			D (D-2) \Gamma(2-\textstyle{{D \over 2}})
			(\Gamma(\textstyle{{D \over 2}}))^3
		}
\eeq
 is the critical value of $N$
 which corresponds to $N_{\rm cr}(\infty)$
 appearing in the subsection \ref{sec:Proof_of_NontrivialSolution},
 as will be shown later.

The solution of Eq.(\ref{eqn:Diff_Eq_case1})
 with the IR B.C. as well as the normalization condition
 has different form for each value of $N$:
\renewcommand{\arraystretch}{2}
\begin{itemize}
	\item Region I ($N < \Ncr$):
	\beq
		\begin{array}{rcl}
			\Sigma(x)
				&=&
			\displaystyle{
				{
					\sigma_m^{{1 \over D-2}}
				\over
					\sin \left( {\omega \over 2} \delta \right)
				}
				\left(
					{\sigma_m \over x}
				\right)^{\frac12} \sin 
				\left\{
					{\omega \over 2}
					\left[
						\ln {x \over \sigma_m} + \delta
					\right]
				\right\},
			}
			\\
			\omega
				&=&
			\displaystyle{
				\sqrt{{\Ncr \over N} - 1}
			}, \;\;\;
			\delta = 2\omega^{-1} \tan^{-1} \omega.
		\end{array}
		\label{eqn:Region_1_case1}
	\eeq

	\item Region ${\rm I\!I}$ ($N = \Ncr$):
	\beq
		\Sigma(x) = \sigma_m^{{1 \over D-2}}
			\left(
				{\sigma_m \over x}
			\right)^{\frac12}
			\left[
				\frac12 \ln {x \over \sigma_m} + 1
			\right].
		\label{eqn:Region_2_case1}
	\eeq

	\item Region ${\rm I\!I\!I}$ ($N > \Ncr$):
	\beq
		\begin{array}{rcl}
			\Sigma(x)
				&=&
			\displaystyle{
				{
					\sigma_m^{{1 \over D-2}}
				\over
					\sinh
					\left(
						{\omega^\prime \over 2} \delta^\prime
					\right)
				}
				\left(
					{\sigma_m \over x}
				\right)^{\frac12} \sinh
				\left\{
					{\omega^\prime \over 2}
					\left[
						\ln {x \over \sigma_m} + \delta^\prime
					\right]
				\right\},
			}
			\\
			\omega^\prime
				&=&
			\displaystyle{
				\sqrt{1 - {\Ncr \over N}}
			}, \;\;\;
			\delta^\prime
				=
			2\omega^{\prime -1} \tanh^{-1}\omega^\prime.
		\end{array}
		\label{eqn:Region_3_case1}
	\eeq
\end{itemize}
\renewcommand{\arraystretch}{1}
Since only the solution of Region I can satisfy the UV B.C.
 for $\sigma_m \neq 0$,
 we conclude that the dynamical mass is generated
 only in case of $N < \Ncr$,
 which means $\Ncr$ is actually the critical value
 separating the symmetric phase and the broken one.
As already mentioned in our previous paper,\cite{IKSY} \ 
 in (2+1) dimensions the critical value $\Ncr = 128/3\pi^2$
 is identical to the one in QED$_3$.\cite{KEIT} \ 
Furthermore,
 the UV B.C. Eq.(\ref{eqn:UVBC_case1}) gives the relation
 between $N$ and $\sigma_m = (m/\Lambda)^{(D-2)}$ as
\beq
	{\omega \over 2}
		\left[
			\ln {1 \over \sigma_m} + 2\delta
		\right] = n\pi,\;\; n = 1,2,\cdots \;\;,
	\label{eqn:UVBCresult_case1}
\eeq
 where the solution with $n = 1$
 corresponds to the nodeless (ground state) solution
 whose scaling behavior is read from Eq.(\ref{eqn:UVBCresult_case1}):
\beq
	{m \over \Lambda}
		=
	e^{{2\delta \over D-2}} \exp
	\left[
		-{2\pi \over (D-2)\sqrt{\Ncr/N-1}}
	\right].
	\label{eqn:ScalingLaw_case1}
\eeq
As was done in Refs.~\citen{Mir,IKSY},
 we recognize it as determining
 the cut-off dependence of $N = N(\Lambda)$,
 by which we can have a finite $m$
 when $\Lambda$ is taken to be infinite (continuum limit).
In fact, if $N(\Lambda)$ goes to $\Ncr$ continuously
 when $\Lambda \rightarrow \infty$,
 it is possible to keep $m$ finite,
 and hence the beta function
 which determines the behavior of $N(\Lambda)$
 with respect to $\Lambda$ is derived as
\beq
	\beta_\Lambda(N)
		\equiv
	\Lambda {\del N(\Lambda) \over \del \Lambda}
		=
	{(D-2) \over \pi \Ncr}
	N^2
	\left(
		{\Ncr \over N} - 1
	\right)^{\frac32},
	\label{eqn:Beta_case1}
\eeq
 where this form is valid for $N$ being nearly below $\Ncr$.
Eq.(\ref{eqn:Beta_case1}) implies that
 the critical value $\Ncr$ is a UV fixed point
 on which the continuum theory will be determined.
Furthermore the anomalous dimension
 of the operator $\br{\psi}\psi$,
 which is relatively large ($\AD \sim O(1)$)
 in models inducing the dynamical mass generation,
 is obtained by
\bea
	\AD(N)
		&=&
	{
		d \ln \la \br{\psi} \psi \ra_\Lambda
	\over
		d \ln \Lambda
	} \nonumber \\
		&=&
	{D-2 \over 2} + 2N\beta_\Lambda(N)
	\nonumber \\
		&=&
	{D-2 \over 2}
	\left[
		1 - {4N^3 \over \pi\Ncr}
		\left(
			{\Ncr \over N} - 1
		\right)^{\frac32}
	\right].
	\label{eqn:AD_case1}
\eea
In Eq.(\ref{eqn:AD_case1})
 the vacuum expectation value $\VEVofMass$
 is given as
\beq
	\VEVofMass
		=
	-
	{
		32
	\over
		(4\pi)^{{D \over 2}} (D-2)
		\Gamma(\textstyle{{D \over 2}}) \Ncr
	}
	\cdot
	N^2
	\cdot
	m^{{D \over 2}} \Lambda^{{D-2 \over 2}},
	\label{eqn:VEV_case1}
\eeq
 where we used the following expression
 for $\VEVofMass$ in the bifurcation theory:
\bea
	\VEVofMass
		&=&
	-
	{
		2 \tr I \; N
	\over
		(4\pi)^{{D \over 2}}
		\Gamma(\textstyle{{D \over 2}})
	}
	\int^\Lambda_0 \! dp \; p^{D-1}
	{
		B(p)
	\over
		p^2 + B^2(p)
	} \nonumber \\
		&\simeq&
	-
	{
		2 \tr I \; N
	\over
		(4\pi)^{{D \over 2}}
		\Gamma(\textstyle{{D \over 2}})
	}
	\int^\Lambda_m \! dp \; p^{D-3} B(p),
	\label{eqn:VEV}
\eea
 and substituted $B(p)$ into the solution in Region I.
As is expected,
 Eq.(\ref{eqn:AD_case1}) tells us that
 the anomalous dimension $\AD(N)$
 is actually large, especially $\AD(N) \simeq (D-2)/2$
 in the vicinity of the critical point.

The above renormalization group functions are obtained
 along with the horizontal line in $(N,1/g)$ plane,
 which is specified by $1/g \equiv 0$.
However, it might not be natural
 to give the cut-off dependence to $N$
 which should be an integer number.
Thus we hope to know the solution for a finite $g$
 to attach the $\Lambda$-dependence not to $N$ but to $g$,
 although we cannot solve the SD equation exactly for any $g$.
In the rest of this section
 we will solve the SD equation Eq.(\ref{eqn:Bifurcation})
 for finite $g$ though restricted region.

\subsection{Analytical Solution {\rm :} $g \gg 1$ case}
\label{sec:Anal_Sol_2}

First we attempt to find
 the next-to-leading order solution in $1/g$ expansion,
 which means that we expands Eq.(\ref{eqn:Bifurcation})
 in $1/g$ up to $O(1/g)$.
In this expansion
 the integrand of Eq.(\ref{eqn:SDKernel}) is reduced to
\bea
	&&
	\eta(k^2)
		=
	{
		(D-2)^2
	\over
		D
	}
	\left\{
		1 - (D-1) C_D G^{-1} k^{2-D}
	\right\}
	+ O(G^{-2}),
	\label{eqn:Eta_Expansion}\\
	&&
	d(k^2)[D - \eta(k^2)]
		=
	{
		4(D-1) C_D
	\over
		D
	}
	\cdot
	{
		1
	\over
		k^{D-2} + f_D/G
	}
	+ O(G^{-2}),
	\label{eqn:Kernel_Expansion}
\eea
 where
\beq
	f_D \equiv {(4-D)D \over 4} C_D
\eeq
 is a value of $O(1)$ in $\TwoDFour$.
Eq.(\ref{eqn:Kernel_Expansion}) seems to have a simple form,
 but we use the ansatz
\beq
	k^2 = p^2 + q^2 - 2pq \cos \theta
		\simeq
	\max
	\left(
		p^2, q^2
	\right)
	\label{eqn:Anzatz}
\eeq
 for Eq.(\ref{eqn:KernelArg})
 to calculate the integral kernel analytically.
This ansatz will be checked by the numerical study
 in \S\ref{sec:Num_Sol} even in (2+1) dimensions.
By substituting Eq.(\ref{eqn:Anzatz})
 into Eq.(\ref{eqn:Kernel_Expansion}),
 we can evaluate Eq.(\ref{eqn:Form_Kernel}) as
\beq
	K(p,q;G)
		=
	K_D
	\left[
		\theta(p^{D-2}-q^{D-2}){1 \over p^{D-2} + f_D/G}
			+
		(p \leftrightarrow q)
	\right],
	\label{eqn:Kernel_case2}\\
\eeq
 where $K_D$ is defined by Eq.(\ref{eqn:Kernel_Coefficient}).
Similarly to the previous subsection,
 by rescaling with $\Lambda$ and translating
 into the differential equation,
 we obtain
\beq
	{d \over dx}
	\left(
		(x+f_D/g)^2 {d \Sigma(x) \over dx}
	\right)
		+
	{\Ncr \over 4N} \Sigma(x)
		=
	0,
	\label{eqn:Diff_Eq_case2}
\eeq
\vspace{-1em}
\begin{subeqnarray}
	&\hbox to 10em{$\Sigma^\prime(\sigma_m) = 0,$\hfil}&
	\mbox{(IR B.C.)}
	\slabel{eqn:IRBC_case2} \\
	&\hbox to 10em{$%
	\left[
		(x + f_D/g) \Sigma
	\right]^\prime_{x=1} = 0,$\hfil}&
	\mbox{(UV B.C.)}
	\slabel{eqn:UVBC_case2}
	\label{eqn:BoundaryCondition_case2}
\end{subeqnarray}
 with the normalization condition
 being the same as Eq.(\ref{eqn:Normalization_case1}).
The solution of Eq.(\ref{eqn:Diff_Eq_case2}) is easily
 obtained by substituting $x$ into $x + f_D/g$
 in the solution of $g \rightarrow \infty$ case,
 and the explicit form of the solution is given by
\beq
	\Sigma(x)
		=
	{
		\sigma_m^{{1 \over D-2}}
	\over
		\sin
		\left(
			{\omega \over 2} \delta
		\right)
	}
	\left(
		{\sigma_m + f_D/g \over x + f_D/g}
	\right)^{\frac12} \sin
	\left\{
		{\omega \over 2}
		\left[
			\ln{x + f_D/g \over \sigma_m + f_D/g} + \delta
		\right]
	\right\}.
	\label{eqn:Solution_case2}
\eeq
In Eq.(\ref{eqn:Solution_case2}) $\omega$ and $\delta$ are
 the same as those defined in Eq.(\ref{eqn:Region_1_case1}),
 and we wrote down only the solution
 corresponding to the case of $N < \Ncr$,
 because only this solution has a possibility
 to satisfy the UV boundary condition Eq.(\ref{eqn:UVBC_case2}).
The UV boundary condition tells us the scaling relation
\beq
	{m \over \Lambda}
		=
	\left\{
		(1+f_D/g) e^{2 \delta - {2 \pi \over \omega}} - f_D/g
	\right\}^{{1 \over D-2}}
		=
	(1 + g_{\rm cr}/f_D)^{{-1 \over D-2}}
	\left[
		1 - {g_{\rm cr} \over g}
	\right]^{{1 \over D-2}},
	\label{eqn:ScalingLaw_case2}
\eeq
 where,
\bea
	g_{\rm cr}
		\equiv
	f_D
	\left( e^{{2\pi \over \omega} - 2\delta} - 1 \right)
	\label{eqn:Critical_case2}
\eea
 is the critical coupling
 above which the dynamical mass is generated.
{}From Eq.(\ref{eqn:ScalingLaw_case2}) the beta function of $g$
 can be seen as follows:
\beq
	\beta_\Lambda(g)
		=
	\Lambda {\del g \over \del \Lambda}
		=
	(D-2) g \left( 1 - {g \over g_{\rm cr}} \right),
	\label{eqn:Beta_case2}
\eeq
 which coincides with the one of the $D$-dimensional GN model
{}\cite{KY,KTY}
 except for the expression of $g_{\rm cr}$.
At this moment it is not clear
 why these are the same to each other
 in contrast to the case of $g \ll 1$
 to be discussed in the next subsection.
Furthermore the anomalous dimension is given as
\bea
	\AD(g)
		&=&
	(D-2)
	{
		1 + f_D/g_{\rm cr}
	\over
		1 + f_D/g
	},
	\label{eqn:AnomalousDimension_case2}
\eea
 which is calculated by using
\beq
	\VEVofMass
		=
	-
	{
		32 N^2
	\over
		(4\pi)^{{D \over 2}} (D-2)
		\Gamma(\textstyle{{D \over 2}}) \Ncr
	}
	\cdot
	(\sigma_m + f_D/g) m \Lambda^{D-2}
	\label{eqn:VEV_case2}
\eeq
 obtained by substituting
 Eq.(\ref{eqn:Solution_case2}) into Eq.(\ref{eqn:VEV}).
Again $\AD(g \simeq g_{\rm cr}) \simeq D-2$
 in coincidence with GN model in $D$ dimensions.\cite{KY,KTY}

\subsection{Analytical solution {\rm :} $g \ll 1$ case}
\label{sec:Anal_Sol_3}
We finally analyze the case
 that the four fermion coupling constant is very small.
In this case the integral defining the kernel function
 of the SD equation can easily be performed explicitly as
\bea
	d(k^2)[D - \eta(k^2)]
		&=&
	G + O(G^2),
	\\
	K(p,q;G)
		&=&
	{
		2 D G
	\over
		(4\pi)^{D/2}(D-2)\Gamma(D/2)
	} \equiv M_D G,
	\label{eqn:SDKernel_case3}
\eea
 which reduces a momentum independent soluition.
Then the scaling relation is obtained as
\beq
	{m \over \Lambda}
		=
	\left(
		1 - {g_{\rm cr} \over g}
	\right)^{1/(D-2)},
	\label{eqn:ScalingLaw_case3}
\eeq
 where
\beq
	g_{\rm cr} = {N \over M_D}
		=
	{
		(4\pi)^{D/2} (D-2) \Gamma(D/2) N
	\over
		2 D
	}
\eeq
 is the critical coupling in this case,
 and the beta function and the anomalous dimension
 are obtained as 
\bea
	\beta_\Lambda(g)
		&=&
	(D-2) g \left( 1 - {g \over g_{\rm cr}} \right),
	\label{eqn:Beta_case3}
	\\
	\AD(g)
		&=&
	(D-2) {g \over g_{\rm cr}},
	\label{eqn:AnomalousDimension_case3}
\eea
 respectively, similarly to
 that of the $D$-dimensional GN model.\cite{KY,KTY}

As a summary of this section
 let us notice that
 both $g \gg 1$ and $g \ll 1$ cases
 give the same result $\AD \simeq (D-2)$
 at $g \simeq g_{\rm cr}$,
 which is expected from the fact that
 the anomalous dimension near the critical line
 is a kind of the critical exponents.
On the other hand,
 the result obtained in the infinite limit
 of the coupling constant
 is different from the others,
 since in this case the renormalization group functions
 are calculated along with the $N$-axis in $(N,1/g)$ plane.
The similar situation can be seen
 in the gauged NJL model.\cite{KTY}

\section{Numerical Study for the SD equation in (2+1) Dimensions}
\label{sec:Num_Sol}
As indicated in the last section,
 we check the validity of the ansatz
 used to solve the SD equation for $g \gg 1$.

In this paper we would like to
 show the result only in (2+1) dimensions,
 in which the integral kernel can be calculated
 for any value of $g$ as
\beq
	\wt{K}(x,y;g)
		=
	{8 \over \pi^2 x y}
	\left[
		\kappa(x+y;g) - \kappa(|x-y|;g)
	\right],
	\label{eqn:Kernel_D3}
\eeq
 where,
\beq
	\kappa(z;g)
		\equiv
	{1 \over g}
	\left[
		{4gz \over 3} + {1 \over gz}
		-
		\left(
			1 + {1 \over g^2 z^2}
		\right)
		\ln ( 1 + gz )
	\right].
	\label{eqn:Mfunc}
\eeq
In the above equations, we already used
 variables normalized by the cut-off,
 which are used frequently in previous sections.
Since the kernel is solved analytically,
 we can skip a few steps
 in the numerical calculation,
 which include complicated numerical integrations.
Moreover,
 let us restrict to the linearized SD equation
 Eq.(\ref{eqn:Bifurcation})
 instead of the original nonlinear integral equation,
 since the linear integral equation
 can be treated as an eigenvalue problem
 which is easier to treat on computational analysis.
As mentioned in \S{}\ref{sec:Anal_Sol},
 it is enough for investigating
 the structure of the critical line.

The figures following below
 show the result of the numerical calculation
 with the analytical solutions given in the previous section.\par
\vspace{-2em} \par
\begin{wrapfigure}{c}{\textwidth}
	\centerline{%
		\hbox{%
			\epsfxsize=6.6cm
			\epsfbox{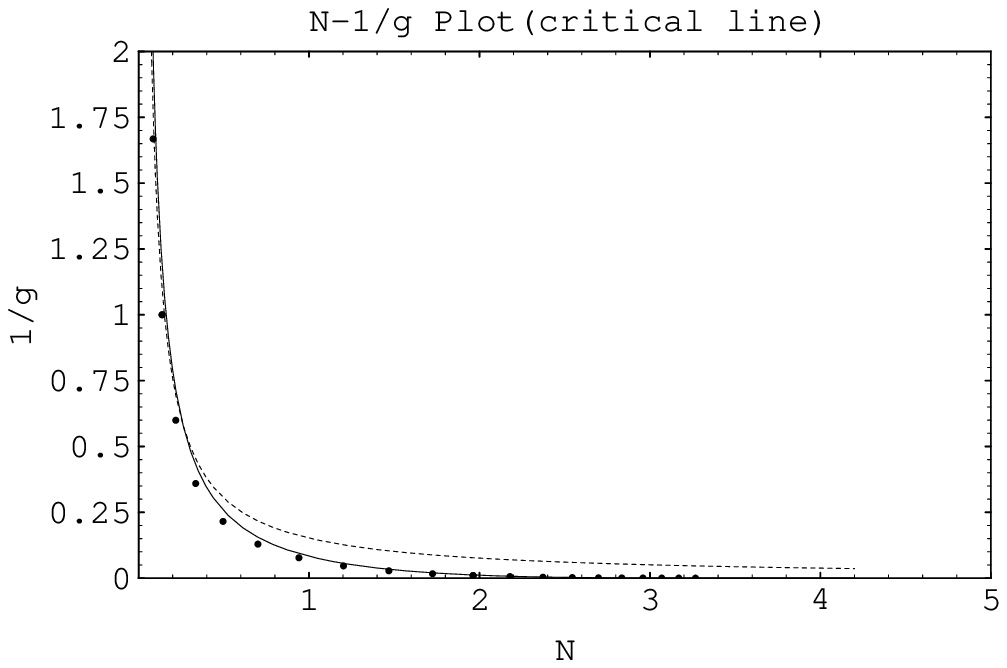}
			\hspace{2em}%
			\epsfxsize=6.6cm
			\epsfbox{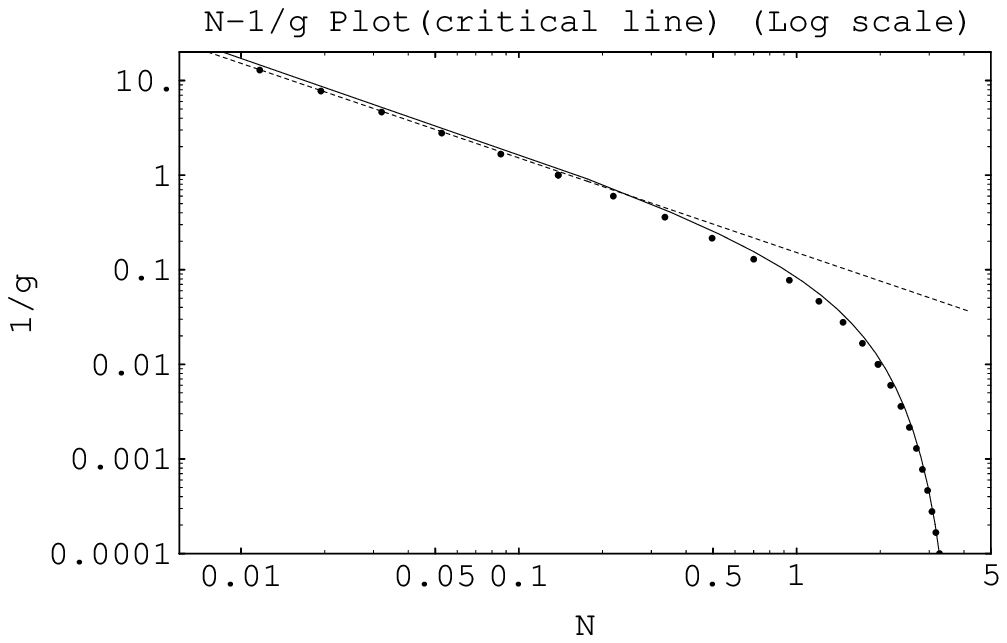}
		}
	}
	\vspace{-3em}
	\caption{The critical line in (2+1) dimensions}
	\label{fig:CriticalLine_D3}
\end{wrapfigure}

The prot points in Fig.\ref{fig:CriticalLine_D3}
 are the result of numerical calculation
 with the kernel of Eq.(\ref{eqn:Kernel_D3}).
Lines represent the analytical solutions,
 a solid one is the solution of $g \gg 1$
 given in \S\ref{sec:Anal_Sol_2},
 and a dashed one is that of $g \ll 1$ given
 in \S\ref{sec:Anal_Sol_3}.

{}From these figures
 it is concluded that
 that the analytical calculation with the ansatz
 well reproduces the critical line
 obtained by the numerical analysis.

\section{Conclusion and Discussion}

As an extension of our previous work,\cite{IKSY} \ 
 we have further studied dynamical fermion mass generation
 of the Thirring model as a gauge theory
 in $D$($\TwoDFour$) dimensions.

After the introduction of the hidden local symmetry
 and the Schwinger-Dyson equation,
 we have completed the existence proof
 of a nontrivial solution of the SD equation
 made in Ref.~\citen{IKSY}
 where an assumption on the property of
 eigenvalue of the integral equation
 was implicitly made.
Furthermore we have shown that
 the critical line in $(N,1/g)$ plane starts from $(0,\infty)$
 and ends at $(N_{\rm cr}(\infty),0)$,
 which can be proven
 only by using the asymptotic behavior of the kernel
 with respect to $g$,
 the dimensionless four-fermion coupling constant.

As natural but nontrivial extensions
 of the previous work,
 we have found the analytical solutions
 of the linearized SD equation (bifurcation theory)
 in $D$($\TwoDFour$) dimensions for some region of $g$.

In the case of $g \rightarrow \infty$,
 the SD equation in $D$ dimensions can be solved exactly
 and the nontrivial solution have been obtained
 in the region $N < \Ncr(< \infty)$,
 which is consistent with
 the result in (2+1) dimensions.\cite{IKSY}
This general $D$-dimensional form of the solution
 tells us that the critical value $\Ncr$
 which depends only on $D$
 is monotonically decreasing for $\TwoDFour$
 from infinity ($D \rightarrow 2$)
 to zero ($D \rightarrow 4$),
 and reproduces the result in $D=3$ of Ref.~\citen{IKSY}.
This result would imply that in $D=4$
 dynamical mass generation does not occur
 even if we takes $g \rightarrow \infty$,
 while there is only the broken phase in $D=2$.
However, there are subtleties in both dimensions
 concerning the UV/IR divergences
 which we did not take into account
 because of the absence of those in $\TwoDFour$ dimensions.
As indicated in Appendix \ref{App:Bifurcation},
 we should not believe the result
 of the linearized SD equation in $D=2$,
 although the SD equation itself is exact for $D=2$,
 since the IR singularity may break
 the validity of this equation.
In case of $D=4$ the UV divergence
 should be properly regularized
 in the evaluation of the propagator
 of the dynamical gauge boson.
Therefore we should study more carefully in this case.

The analytical solutions have been obtained also for finite $g$
 in the cases of $g \gg 1$ and $g \ll 1$.
These have given explicit form of the critical line
 which indicates a finite critical coupling constant $g_{\rm cr}$
 for $N < \Ncr$ as opposed to the result of Ref.~\citen{HP}
 claims $g_{\rm cr}$ for all $N$.
On the other hand, at least qualitatively,
 our result seems to be consistent with that of Ref.~\citen{Kondo}
 which is obtained also based on the HLS
 but through the inversion method instead of the SD equation.
Furthermore, it should be emphasized that
 the numerical studies{}\cite{Hands,KmKm}
 would indicate that there is a finite critical value of $N$.

Moreover, we have found that
 the solutions obtained at both $g \gg 1$ and $g \ll 1$
 gives the same expression for $\beta_\Lambda(g)$
 and the same value of the anomalous dimension
 on the critical line, i.e.,
 $\gamma_m(g) = (D-2)$ at $g=g_{\rm cr}$
 which is identical to that of
 $D$-dimensional Gross-Neveu model.\cite{KY}

Finally, we have confirmed validity of the ansatz
 used in the analytical calculation through numerical study
 of the linearized SD equation with the exact kernel,
 although limited to (2+1) dimensions.
This result has shown that
 the numerical solution agrees with the analytical one
 in both qualitative and quantitative sense.
However, it may be nontrivial
 whether or not we can extend this agreement
 to other dimensions,
 which should be confirmed by the efforts
 to the more complicated numerical studies.

\section*{Acknowledgements}
We would like to thank K.~Yamawaki and A.~Shibata
 for collaboration at early stage.

\appendix
\section{Brief Summary of The Bifurcation Theory}
\label{App:Bifurcation}

In this Appendix
 we will briefly summarize the bifurcation theory,
 and comment on a possibility of break down
 of the bifurcation theory in (1+1) dimensions.

The bifurcation theory is applied to
 such a following functional equation,
\beq
	\phi(x) = \lambda F[\phi;x),
	\label{eqn:FunctionalEqn}
\eeq
 where, $F[\phi;x)$ in the R.H.S. is a functional of $\phi(x)$
 as well as a function of $x$, and $\lambda$ is a parameter.
Here let us assume that
 one of the solutions of Eq.(\ref{eqn:FunctionalEqn})
 is already known,
 which is denoted as $\phi_0(x;\lambda)$.
Hence we are interested in the solution
 which bifurcates from $\phi_0(x;\lambda)$
 continuously at $\lambda = \lambda_{\rm br}$.

As discussed in Ref.~\citen{Bifurcation},
 the necessary condition
 for the existence of such a bifurcated solution is as follows.
{\it The functional $F[\phi;x)$
 is functionally differentiable in the third order
 with respect to $\phi(x)$
 at $\phi(x) = \phi_0(x;\lambda_{\rm br})$
},
 where the bifurcation point $\lambda_{\rm br}$
 is determined as the eigenvalue
 of the linearized equation for $\delta \phi(x)$:
\beq
	\delta \phi(x)
		=
	\lambda
	\int dy
	\left.
		{
			\delta F[\phi;x)
		\over
			\delta \phi(y)
		}
	\right|_{\phi(x) = \phi_0(x)}
	\delta \phi(y).
	\label{eqn:LinearizedEqn}
\eeq
Thus the bifurcation solution is given
 as $\phi(x) = \phi_0(x;\lambda_{\rm br}) + \delta \phi(x)$
 in the vicinity of $\lambda_{\rm br}$.

We can easily check with the SD equation (\ref{eqn:LadderSDeqn_b})
 that the R.H.S. satisfies the necessary condition.
Moreover,
 we can also confirm that the necessary condition is satisfied
 even if we use the fermion full propagator
 when calculating the dynamical gauge boson's propagator
 and the integral kernel of Eq.(\ref{eqn:LadderSDeqn_b}),
 although the kernel cannot be solved explicitly
 in such a case.

Concerning with the above statement,
 we should comment on the case of $D=2$
 in which the exact solution was found
 to be equivalent to a free theory
 having no dimensionful parameter.
However,
 if we solve Eq.(\ref{eqn:Bifurcation}) in $D=2$ naively,
 a nontrivial solution for $g \ge 0$ exists,
 which is constant for the momentum
 as $\delta B(p^2) \equiv m = \Lambda\exp\{-N(1 + \pi/g)\}$.
It should be emphasized that
 the ladder approximation is not wrong in this case,
 although it has been usually suspected.
The reason is that the set of vertex functions
\beq
	\begin{array}{l}
		S(p) = i/\sla{p},\;\;
		D_{\mu\nu}(k) = (g^{-1} + \pi^{-1})g_{\mu\nu},\;\;
		\Gamma^\mu(p,q) = \gamma^\mu,
		\\
		\mbox{other vertex function} = 0
	\end{array}
\eeq
 is the exact solution of the full series of the SD equations,
 which means that the ladder approximation is {\it exact}
 in (1+1) dimensions,
 if the bifurcation theory is valid.

Do these considerations conclude the dynamical mass generation
 in the (1+1)-dimensional Thirring model?
This confliction will be resolved
 by investigating the integral kernel more in detail.
Thus it might be shown that we are not able
 to apply the bifurcation theory to the case of $D=2$
 because of the non-analyticity of $K(p,q;G)$
 with respect to the mass function $B(p^2)$
 near a trivial solution $B(p^2) \equiv 0$, i.e.,
 $\delta K(p,q;G) / \delta B(p^2) |_{B(p^2) \equiv 0} = \infty$.
This property of the kernel is contrary to the necessary condition
 explained above.
Therefore in (1+1) dimension
 we should not naively conclude
 the existence of a nontrivial solution
 through the bifurcation theory.

\section{Proof that $\lambda(g,\sigma_m)$%
 is a decreasing function of $\sigma_m$}
\label{App:EigenValue}

In \S\ref{sec:SD_eqn}
 we needed to prove that the maximal eigenvalue
 of the integral equation (\ref{eqn:EigenValueProblem})
 is monotonically decreasing function with respect to $\sigma_m$.
For this purpose,
 we will prove more general statement in this Appendix.

Let us consider the following eigenvalue problem:
\beq
	\phi(x) = {1 \over \lambda} \int^b_a dy K(x,y) \phi(y),
	\label{eqn:EigenValueProblem2}
\eeq
 where $K(x,y)$ is a real symmetric integral kernel
 which is finite on the region $a \le x,y \le b$.
We denote eigenvalues and corresponding eigenfunctions
 as $\lambda_n(a)$ and $\phi_n(x;a)$,
 in which $n$ is assigned as
 $\lambda_1(a) > \lambda_2(a) > \ldots$.
The dependence of $a$
 is defined by both Eq.(\ref{eqn:EigenValueProblem2})
 and the normalization condition
\beq
	\int^b_a
		dx \phi_m(x;a) \phi_n(x;a) = \delta_{mn},
	\label{eqn:Normalization}
\eeq
 and hence $\lambda_n(a)$ is expressed as
\beq
	\lambda_n(a)
		=
	\int^b_a \int^b_a
		dxdy \phi_n(x;a) K(x,y) \phi_n(y;a).
	\label{eqn:IntegralExpression}
\eeq
By using this explicit expression of $\lambda_n(a)$,
 we can calculate the derivative of $\lambda_n(a)$
 with respect to $a$ as
\bea
	{d \lambda_n(a) \over d a}
		&=&
	-2 \phi_n(a;a)
		\int^b_a dy K(a,y) \phi_n(y;a)
	\nonumber \\
		&&
	\mbox{\hspace{2em}}
	+ 2 \int^b_a \int^b_a
		dxdy {\del \phi_n(x;a) \over \del a} K(x,y)
			\phi_n(y;a)
	\nonumber \\
		&=&
	-2 \lambda_n(a)
	\left(
		\phi_n(a;a)
	\right)^2
		+
	2 \lambda_n(a)
		\int^b_a dx {\del \phi_n(x;a) \over \del a}
			\phi_n(x;a),
	\label{eqn:Derivative1}
\eea
 where, we use the symmetry property of $K(x,y)$.
The second term of the last line of Eq.(\ref{eqn:Derivative1})
 is derived by the differentiation
 of Eq.(\ref{eqn:Normalization}):
\bea
	0
		&=&
	{d \over d a}
		\int^b_a dx \phi_n(x;a) \phi_n(x;a)
	\nonumber \\
		&=&
		-
	\left(
		\phi_n(a;a)
	\right)^2
		+
	2 \int^b_a dx {\del \phi_n(x;a) \over \del a}
		\phi_n(x;a).
	\label{eqn:Derivative2}
\eea
Substituting Eq.(\ref{eqn:Derivative2})
 into Eq.(\ref{eqn:Derivative1}),
 we finally obtain the formula as
\beq
	{d \lambda_n(a) \over d a}
		=
	- \lambda_n(a)
	\left(
		\phi_n(a;a)
	\right)^2\;\;
	\left\{
		\begin{array}{ll}
			\le 0	&	(\lambda_n(a) \ge 0)	\\
			> 0		&	(\lambda_n(a) < 0)
		\end{array}
	\right.
	,
\eeq
 which is the statement that should be proven.

Combining with the fact that
 the maximal eigenvalue $\lambda_1(a)$ has always positive sign,
 we have actually proven that
 $\lambda_1(a)$ is monotonically decreasing function
 for the lower bound of the integral.


\end{document}